# Multi-Site Infant Brain Segmentation Algorithms: The iSeg-2019 Challenge

Yue Sun, Kun Gao, Zhengwang Wu, Zhihao Lei, Ying Wei, Jun Ma, Xiaoping Yang, Xue Feng, Li Zhao, Trung Le Phan, Jitae Shin, Tao Zhong, Yu Zhang, Lequan Yu, Caizi Li, Ramesh Basnet, M. Omair Ahmad, M.N.S. Swamy, Wenao Ma, Qi Dou, Toan Duc Bui, Camilo Bermudez Noguera, Bennett Landman, *Senior Member, IEEE,* Ian H. Gotlib, Kathryn L. Humphreys, Sarah Shultz, Longchuan Li, Sijie Niu, Weili Lin, Valerie Jewells, Gang Li, *Senior Member, IEEE*, Dinggang Shen, *Fellow, IEEE*, Li Wang, *Senior Member, IEEE*

***Abstract*—To better understand early brain growth patterns in health and disorder, it is critical to accurately segment infant brain magnetic resonance (MR) images into white matter (WM), gray matter (GM), and cerebrospinal fluid (CSF). Deep learning-based methods have achieved state-of-the-art performance; however, one of major limitations is that the learning-based methods may suffer from the *multi-site issue,* that is, the models trained on a dataset from one site may not be applicable to the datasets acquired from other sites with different imaging protocols/scanners. To promote methodological development in the community, iSeg-2019 challenge (http://iseg2019.web.unc.edu) provides a set of 6-month infant subjects from multiple sites with different protocols/scanners for the participating methods. Training/validation subjects are from UNC (MAP) and testing subjects are from UNC/UMN (BCP), Stanford University, and Emory University. By the time of writing, there are 30 automatic segmentation methods participating in iSeg-2019. We review the 8 top-ranked teams by detailing their pipelines/implementations, presenting experimental results and evaluating performance in terms of the whole brain, regions of interest, and gyral landmark curves. We also discuss their limitations and possible future directions for the *multi-site issue*. We hope that the multi-site dataset in iSeg-2019 and this review article will attract more researchers on the multi-site issue.**

## I. Introduction

SEGMENTING infant brain images into different tissues, e.g., white matter (WM), gray matter (GM), and cerebrospinal fluid (CSF), is a vital step in the study of early brain development. Due to inherent myelination and maturation process, infant brain images exhibit extremely low tissue contrast, especially at around 6-months old. In iSeg-2017 [1], we have reviewed state-of-the-art methods to deal with the challenge from the extremely low tissue contrast and found that deep learning based methods have shown its promising advancement achieving state-of-the-art performances. However, data from multiple sites, poses a number of challenges for the learning-based segmentation methods. Normally, the trained model based on a dataset from one site may handle the testing images from the same site very well, however, the model often performs poorly on datasets from other sites with different imaging protocols/scanners, which is called the *multi-site issue*. Factors include equipment manufacturer, magnetic field strength, and acquisition protocol, which can affect image contrast/pattern and intensity distribution. One example is shown in Fig. 1 that learning-based methods achieved high accuracy on the validation dataset which is from the same site with the training dataset. Unfortunately, it is not replicated when the testing datasets are from multiple scanners with different protocols/scanners (Table I), where there is a large drop in performance in terms of Dice

Y. Sun, K. Gao, and L. Wang were supported by NIH grants MH109773 and MH117943. Z. Wu, T. Bui, and G. Li were supported in part by NIH grants MH117943. I. H. Gotlib was supported by NIH Grants R21HD090493 and R21MH111978. S. Shultz was supported by NIH grant K01 MH108741. S. Shultz and L. Li were supported by NIH grants P50MH10029, R01EB027147, R01MH119251, and R01MH118534. *(Y. Sun, K. Gao, and Z. Wu are co-first authors.) (Corresponding author: L. Wang, li_wang@med.unc.edu)*

Y. Sun, K. Gao, Z. Wu, T. Bui, W. Lin, V. Jewells, G. Li, and L. Wang are with the Department of Radiology and Biomedical Research Imaging Center, University of North Carolina at Chapel Hill, NC 27599, USA.

Z. Lei and Y. Wei are with Northeastern University, Shenyang, China.

J. Ma is with the Department of Mathematics, Nanjing University of Science and Technology, Nanjing, China.

X. Yang is with the Department of Mathematics, Nanjing University, China.

X. Feng is with the Department of Biomedical Engineering, University of Virginia, Charlottesville, VA, USA.

L. Zhao is with the Diagnostic Imaging and Radiology Department, Children's National Medical Center, Washington, DC, USA.

T. L. Phan and J. Shin are with the Department of Electrical and Computer Engineering, Sungkyunkwan University, Suwon, 16419, Korea.

T. Zhong and Y. Zhang are with the School of Biomedical Engineering, Southern Medical University, Guangzhou 510515, China.

L. Yu is with The Chinese University of Hong Kong, China.

C. Li is with Shenzhen Institutes of Advanced Technology, China.

R. Basnet, M. Ahmad, M.N.S. Swamy are with the Department of Electrical and Computer Engineering, Concordia University, Montreal, Canada.

W. Ma is with the School of Informatics, Xiamen University, China.

Q. Dou is with the Department of Computer Science and Engineering, The Chinese University of Hong Kong, China.

C. Noguera and B. Landman are with Electrical Engineering and Computer Science, Vanderbilt University, Nashville, TN 37204, USA.

I. H. Gotlib is with the Department of Psychology, Stanford University, California 94305, USA.

K. L. Humphreys is with the Department of Psychology and Human Development, Vanderbilt University, Nashville, TN 37204, USA.

S. Shultz and L. Li are with the Emory University, Atlanta, GA 30322, USA.

S. Niu is with the School of Information Science and Engineering, University of Jinan, Jinan 250022, China.

D. Shen is with the Department of Research and Development, Shanghai United Imaging Intelligence Co., Ltd., Shanghai, China, and also Department of Brain and Cognitive Engineering, Korea University, Seoul 02841, Republic of Korea.

Coefficient (DICE) metric (WM: from 0.90 to 0.86; GM: from 0.92 to 0.82; CSF: from 0.92 to 0.83). The *multi-site issue* hinders the popularity of learning-based methods. Currently, few investigators have addressed the multi-site issue with the following approaches: few-shot learning [2], domain adaption [3], transfer or distributed transfer learning [4], and adversarial learning [5]. These existing methods still require either a small number of labels (annotations) from other sites for fine-tuning or a large number of images from other sites for adaption. However, the annotation from other sites is prohibitively time-consuming and expensive, and usually only a small number of images are acquired in a pilot study before real-world healthcare applications are carried out.

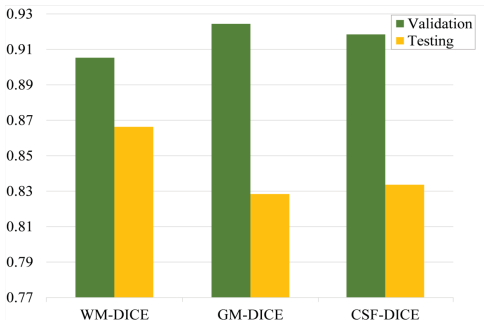

Fig. 1. Multi-site issue: learning-based models often perform well on the validation dataset from the same site with the training dataset, whereas perform poorly on other sites with different imaging protocols/scanners, with large performance decreasing. The validation dataset and testing dataset are provided by iSeg-2019 (the higher the DICE, the better the performance), where WM/GM/CSF-DICE denotes the segmentation performance on WM/GM/CSF in terms of DICE.

To attract more researchers to address the multi-site issue, in 2019, we organized a MICCAI grand challenge on 6-month infant brain MRI segmentation from multiple sites: iSeg-2019, http://iseg2019.web.unc.edu. Although a naïve way is to train models based on multiple sites with different imaging protocols/scanners, the combinations of protocols/scanners are infinite. There is no way to include every possible case. Therefore, the goal of iSeg-2019 is to promote to training models based on one site to fit for all sites. Note that iSeg-2019 is a follow-up challenge of iSeg-2017 [1] in which the training and testing subjects from the same site. In iSeg-2019, the training subjects are randomly chosen from Multi-visit Advanced Pediatric (MAP) Brain Imaging Study and testing subjects are from three sites: University of North Carolina at Chapel Hill/University of Minnesota (Baby Connectome Project, BCP), Stanford University, and Emory University. It is worth noting that the imaging parameters/scanners from the three sites are different from the training dataset, with their imaging protocols/scanners listed in Table I. Labels from experts are only available for training images. Researchers are invited to submit their automatic segmentation results on the validation and testing images. At the time of writing this paper, 30 teams have submitted their segmentation results on the iSeg-2019 website. Section II provides the cohort used for this challenge, following with the details of the 8 top-ranked methods in Section III. After that, Section IV and V elaborate on the performance, limitations and possible future directions, and Section VI concludes the challenge.

## II. Multi-Site Dataset

Since the iSeg-2019 challenge aimed to promote automatic segmentation algorithms on 6-month infant brain MRI from multiple sites, we selected MR images from four different sites as training, validation and testing datasets respectively. Note that the training images and validation images are from the same site. To ensure the consistency of all infant brain images and alleviate bias effects caused by non-algorithmic factors, we employed the same selection criteria and same preprocessing. Specifically, we selected normal brain MRIs with an average age of 6.0±0.8 months to ensure the data consistency. Second, the resolution of all images was resampled into 1.0×1.0×1.0 $mm^3$. Third, the same standard imaging preprocessing steps were performed, including skull stripping [6], intensity inhomogeneity correction [7], and removal of the cerebellum and brain stem. Finally, each preprocessed image was examined and errors were manually corrected by experts (Dr. Li Wang and Dr. Valerie Jewells).

TABLE I
Dataset Information of the iSeg-2019 Challenge

| | Site | Scanner | Modality | TR/TE (ms) | Resolution ($mm^3$) | Number |
|---|---|---|---|---|---|---|
| Training | UNC (MAP) | Siemens (3T) | T1w | 1900/4.4 | 1.0×1.0×1.0 | 10 |
| Validation | | | T2w | 7380/119 | 1.25×1.25×1.95 | 13 |
| Testing | UNC/UMN (BCP) | Siemens (3T) | T1w | 2400/2.2 | 0.8×0.8×0.8 | 6 |
| | | | T2w | 3200/564 | 0.8×0.8×0.8 | |
| | Stanford University | GE (3T) | T1w | 7.6/2.9 | 0.9×0.9×0.8 | 5 |
| | | | T2w | 2502/91.4 | 1.0×1.0×0.8 | |
| | Emory University | Siemens (3T) | T1w | 2400/2.2 | 1.0×1.0×1.0 | 5 |
| | | | T2w | 3200/561 | 1.0×1.0×1.0 | |

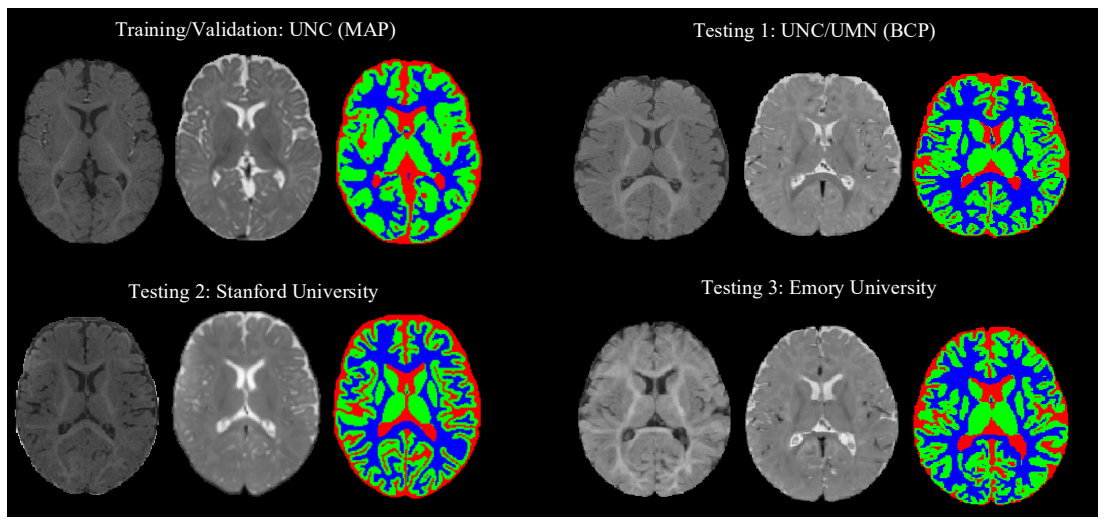

(a)

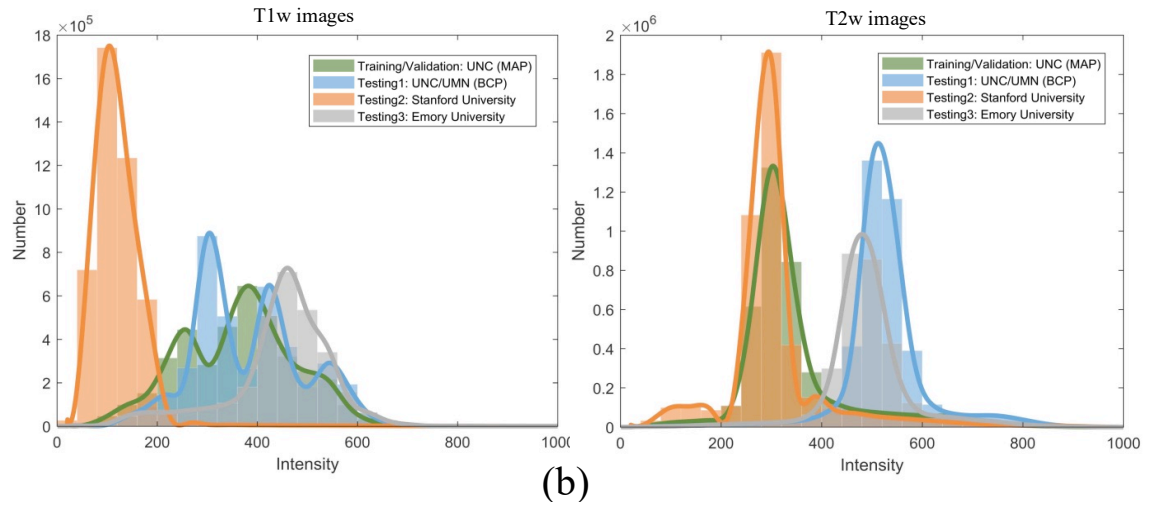

(b)

Fig. 2. T1w and T2w MR images of infant subjects scanned at 6 months of age (isointense phase) from four sites with different imaging protocols/scanners, i.e., UNC (MAP), UNC/UMN (BCP), Stanford University, and Emory University, provided by iSeg-2019. (a) Intensity images and the corresponding ground truth from each site. From left to right: T1w MR image, T2w MR image, and segmentations. (b) Average intensity distribution of T1w and T2w images from four sites marked by different colors.

For training and validation, MR scans were chosen from Multi-visit Advanced Pediatric (MAP) Brain Imaging Study. For testing, MR scans were randomly chosen from three sites, i.e., UNC/UMN (BCP), Stanford University, and Emory University. Detailed imaging protocols and scanners about datasets are listed in Table I. All sites utilized Siemens scanners except for Stanford University which utilized a GE scanner. Furthermore, Fig. 2(a) shows T1w, T2w and manual annotations of WM, GM and CSF of 6-month infant subjects from four sites, and Fig. 2(b) plots average intensity distributions of T1w and T2w images from four sites. It can be clearly seen that there are large differences in the intensity distributions and histogram shape for the T1w images from Stanford University compared to the other sites. These differences cause significant challenges for learning-based methods.

Reliable manual segmentations are generated for training and quantitative comparisons. For the 6-month subjects with follow up scans, we used a longitudinal guided segmentation algorithm [8] to first generate an initial segmentation. For the cases without follow up scans, we used an anatomy-guided densely-connected U-Net [9] to generate an initial segmentation. The initial segmentations were later followed by manual editing. Details of manual protocol are available in [10]. To maximally alleviate the potential bias from the automatic segmentations, we have spent many efforts (20±5 hours) on the manual correction for each subject, with 200,000±10,500 voxels (25%±1.3% of total brain volume) re-corrected. As shown in Fig. 2(a), brain MRIs were manually segmented into 3 classes, i.e., WM, GM and CSF, where myelinated and unmyelinated WM is marked blue, cortical and subcortical GM is marked green and extracerebral/intraventricular CSF is marked red. Finally, we provided 10 infant subjects for training, 13 infant subjects for validation, and 16 infant subjects from three different sites for testing, as detailed in Table I. Note that the manual annotations of training subjects were provided, together with the T1w and T2w images. While annotations of validation/testing subjects are not provided to the participants. The segmentation results for the validation/testing dataset can be submitted maximally 3 times for evaluation and only the latest/best results were recorded.

## III. Methods And Implementations

For the iSeg-2019 Challenge, a total of 30 teams successfully submitted their results to the website before the official deadline. We describe all participating teams with affiliations and features used in their methods in Appendix Table I. Furthermore, we summarize the performance of all teams in Appendix Table II and find that one team out of 29 did not utilize a deep learning technique. Of the remaining 28 teams using convolutional neural networks, 22 teams adopted the U-Net architecture [9], which is a strong baseline for medical image segmentation. In this section, we will describe the 8 top-ranked methods with the corresponding source codes listed in Appendix Table X.

### A. QL111111: Northeastern University, China

Due to the excellent performance of 3D U-Net in medical image segmentation, Lei *et al*. propose a novel method based on 3D U-Net combined with attention mechanism [11]. In addition, they also use dilated convolution in downsampling, so that the semantic information can be extracted in the encoding phase without a lot of details being lost. To deal with site differences between training images and testing images, T1w and T2w images are randomly cropped to 32×32×32, and they set random contrast to 4.64~4.66 for T1w images, and 1.34~1.36 for T2w images. They further apply gamma correction for T2w images. Then both T1w and T2w images are standardized and sent to the network.

Their proposed method structures are presented in Fig. 3. First residual-network-based structure is used in the downsampling including three DCP blocks. Then they use a 1×1×1 Conv in left side, and the right side contains a dila-block module, an activated convolutional layer with a convolution kernel size of 3×3×3 and a stride of 2. This dila-block consists of four dilated convolutions and each dilated convolution includes a pre-activation layer, then they performed concatenation and pass them to next stage. In the upsampling stage, self-attention is utilized to capture long-range dependencies, that is, to aggregate the information of feature maps. When they process the 3D images with attention mechanism, it is necessary to convert 3D data into 2D vectors. They first represent a $D \times W \times H$ cube of the 3D image as a vector of $D \times W \times H \times C$. Then the data after dimensionality reduction operation are operated with attention block. Different from the spatial attention, channel characteristics are directly calculated from the image which are performed dimensionality reduction operation. The channel map for each high-level feature can be thought of as a class-specific response, with different semantic responses associated with each other. The contribution of their work can be summarized as follows: (1) They utilized the convolution of different dilation rates to effectively capture multi-scale information. (2) The proposed attention method can effectively encode the wider context information and mine the interdependencies between channel maps.

For network training, they use cross-validation to train, taking 9 samples as the training dataset, and the remaining one as the validation. During testing, they input the test image to the cross-trained model, and then perform a majority vote to get the final output images. The optimization of network parameters is performed via Adam optimizer. Learning rate is initialized as 3.3e-3 and they set weight decay as 2e-6. Their proposed method is implemented in Python using TensorFlow framework. Experiments are performed on a computational server with one NVIDIA 1080Ti with 11GB of RAM memory. Each training session takes about 3 hours. It takes about 2 minutes to segment a 3D MRI image on NVIDIA 1080Ti.

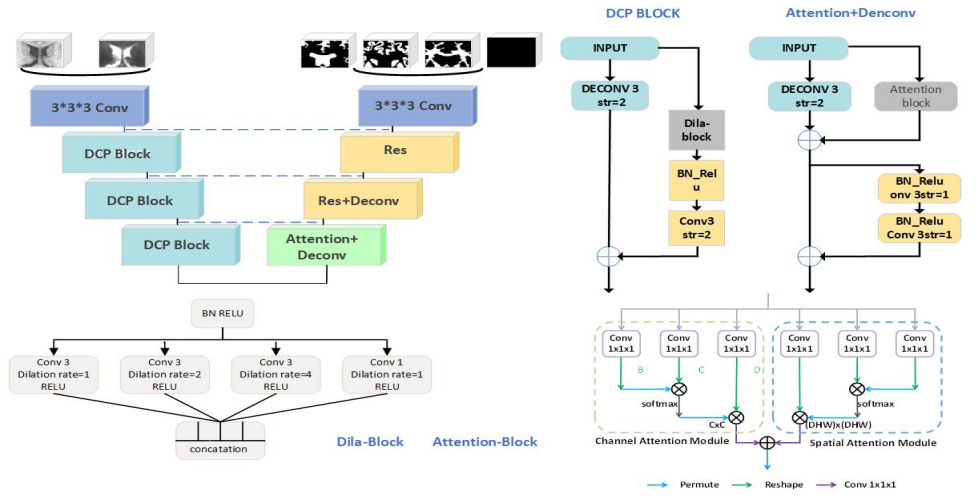

Fig. 3. Team *QL111111*: the structure of 3D U-Net based on attention mechanism.

### B. FightAutism: Nanjing University of Science and Technology, China

To deal with large differences between training images and testing images from multiple sites, Jun *et al*. employed histogram matching to alleviate the intensity variance. Specifically, the intensity distributions of all the testing subjects are adjusted to a randomly selected subject from the validation dataset.

The employed architecture is a naïve 3D U-Net [12] which consists of one down-sampling path and one up-sampling path. Each path includes 5 convolution blocks and each block comprises of 3×3×3 convolution layer, instance normalization layer and leaky rectified linear unit. Long skip connections are also used in the same resolution between down-sampling and up-sampling paths.

Both T1w and T2w images are used to train the U-Net. The pre-processing includes foreground (non-zero regions) cropping and Z-Score normalization. All the 10 training subjects are used for training. Data augmentation includes random rotation, scaling, mirroring, and gamma transformation. The optimizer is Adam with an initial learning rate of 3e-4. During testing, test time augmentation is employed by mirroring along all axes. The implementation is based on PyTorch and nnU-Net [13].

### C. xflz: Children's National Medical Center, USA

Feng *et al*. propose an optimized 3D U-Net with tissue-dependent intensity augmentation for segmentation of 6-month infant brain MRI. To deal with the site differences, they use a tissue-dependent intensity augmentation method to simulate the variations in contrast. The intensity values of each tissue (i.e., WM, GM and CSF) are multiplied by different factors randomly selected during training to emulate the changes of contrast. The 3D U-Net structure with optimization in network structure, training and testing strategies are used to perform the segmentation task.

The network structure is based on 3D U-Net. The encoding path includes three blocks with each block containing two 3×3×3 convolution layers and one 3D max-pooling layer. Two additional convolution layers are added at the bottom. Three decoding blocks are used with long range connection from the corresponding encoding block. Parametric rectifier (PReLU) is used instead of conventional rectifier (ReLU). Compared with the original U-Net for 2D images, the number of features is increased to capture more 3D information, yielding a wider network. As the input of the network, T1w and T2w images are concatenated; the output is the probabilistic maps of WM, GM, CSF and background.

During each iteration of training, after reading the original T1w and T2w images and the label map, the voxel intensities at each region are multiplied by different factors randomly sampled from 0.9 to 1.1 using a uniform distribution. This simulates the effect of different imaging protocols. Randomly sampled Gaussian noise with different standard deviations is then added to the resulting T1w and T2w images, respectively. Furthermore, due to left-right symmetry of the brain, the images are randomly flipped. A patch of 64×64×64×2 is randomly extracted from the original images and fed to the network. The model is trained on the provided data for 4000 epochs and the total time was about 7.5 hours on a NVidia Titan Xp GPU. During deployment, a sliding window approach is used with a stride of 16 and for each voxel, the final probability is obtained by averaging all outputs from overlapping patches. No post-processing is performed. The deployment time is about 1 minute per case.

Although the method achieved good performance in the validation dataset, the performance dropped significantly on the multi-site testing dataset, indicating that the augmentation method is not sufficient to fully emulate the imaging differences. As the imaging parameters including TR and TE are available, they also tried to use a more complex model based on MR physics to simulate the images using different sets of protocols; however, as the MR sequence is very complicated, only using TR, TE, T1w and T2w information and a simple signal decay model fails to yield realistic images. For simplicity and robustness, they just randomly scale the regional intensities differently.

### D. trung: Media System Laboratory, Sungkyunkwan University, South Korea

Trung *et al*. introduce a method for 6-month infant brain MRIs, called Cross-linked FC-DenseNet (cross-linked fully convolution-DenseNet), as shown in Fig. 4. First, they add an end-to-end concurrent spatial and channel squeeze & excitation (scSE) [14], followed by every Dense block, which allows to explore the interdependency among the channels. Second, they combine the 3D FC-DenseNet with the cross-links (such as down-sampling links) [15], which enable to learn more features from the contracting process.

Fig. 4 shows the proposed network with three phases: 3D FC-DenseNet, concurrent scSE, and HyperDenseNet. The network consists of two basic main paths: a contracting/down-sampling path and an expanding/up-sampling path. The initial part of the network has three 3×3×3 convolutions with stride 1 followed a batch normalization layer (BN) and a ReLU that generates 64 output feature maps. The contracting/down-sampling path with three dense blocks with a growth rate of k = 16 [16] is exploited. Each dense block has four BN-ReLU-Conv(1×1×1)-BN-ReLU-Conv (3×3×3). To alleviate the over-fitting after this Conv (3×3×3), they use a dropout layer with a dropout rate of 0.2 [17]. After each dense block, they add scSE block [14] to explore the interdependencies between the channels. After scSE block, the transition block includes BN-ReLU-Conv (1×1×1)-BN-ReLU followed by a convolution layer of stride 2 to reduce feature map resolutions while preserving the spatial information [14]. Meanwhile, for recovering the feature resolution, the expanding/up-sampling path has three convolution layers. To reuse the feature map information of different sizes during the expanding/up-sampling path, the third convolution layer acquires outputs from each transition block in the contracting path using the cross-link path. They resize the outputs of the first and last transition block using max pooling and up-convolution of 2×2×2 by stride 2 to get equal resolution with second transition output and concatenate all of them as input of third convolution layer. Also, the convolution layer uses three different sizes of feature maps: the output of initial part, the output of the first transition block, the output of the third convolution layer. Otherwise, they use two different sizes of feature map in the first convolution layer: the output of the initial part, the output of the second convolution layer. This cross-link path allows the model to capture multiple contextual information from the different layer's features. Furthermore, it improves a gradient flow within a limited dataset. Each convolution layer consists of ReLU-Conv (3×3×3)-BN. Finally, a classifier consisting of ReLU-Conv (1×1×1) is used to classify the concatenation feature maps into target classes.

They implement and train the proposed network within a NVIDIA TitanXP and Pytorch framework. First, they randomly crop sub-volume samples with size 64×64×64 voxels which are extracted in the center of the voxel they are interested in. The network is trained with an Adam optimizer and 6000 epochs with a mini-batch size of 2. The learning rate and weight decay are set to 0.0002 and 0.0006 for the proposed network, respectively. They use 9 subjects for training and 1 subject for validation from the iSeg-2019 training dataset. In the inference phase, the same size in training with stride of 16×16×16 is used. The testing duration is approximately 40 hours for training and 6 minutes for testing each subject.

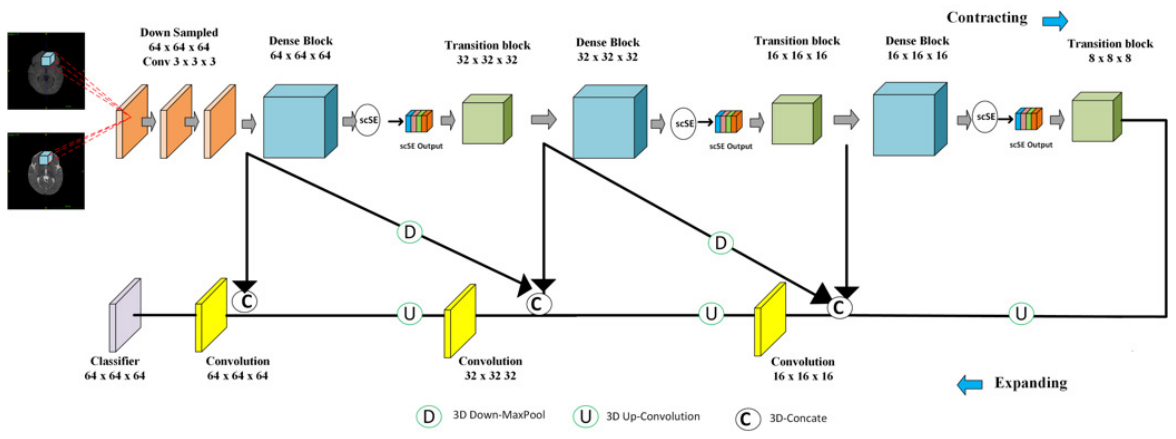

Fig. 4. Team *trung*: proposed cross-linked FC-DenseNet architecture for volumetric segmentation.

### E. Tao_SMU: Southern Medical University, China

Zhong *et al*. propose an attention-guided full-resolution network for segmentation of 6-month infant brain MRIs, which is extended from the full-resolution residual network [18] and attention mechanism [19]. The network consists of a 3D two-stream full-resolution structure and two types of 3D attention modules, including Dual Self-Attention Module and Dual Pooling-Attention Module. First, to address spatial information loss, the 3D-based full-resolution architecture is constructed to preserve high-resolution information by keeping a separate high-resolution processing stream. Second, this architecture is combined with attention mechanism to capture global relationship and generate more discriminative feature representations through spatial and channel axes. To deal with large differences between training images and testing images from multiple sites, they augment the training data with two methods. First, they transform each subject from different dimensions, i.e., transpose each subject by swapping 3 axes in different combination, to generate 5 new subjects, thus providing more features from different views. Second, each training subject would be randomly adjusted contrast and lightness using the parameter ±20%. Model with these data augmentation shows a slight improvement compared with previous models.

Proposed architecture is detailed in Fig. 5. The two-stream structure combines multi-scale context information using two processing streams. The resolution stream carries information at full resolution for precise segmentation of boundaries. The semantic stream acquires high-level features for class identification. In this model, stride size is set as 2 or 4 to achieve magnification/reduction of feature maps 2 or 4 times. Except for the last convolutional layer for classification, all convolution and deconvolution sizes are set as 3×3×3. During the entire process, the repeated multi-scale fusion is realized by exchanging the feature information between the two streams. Meanwhile, two types of attention mechanism are used to improve model feature extraction capabilities. In the bottom block, the fusion features will pass through Dual Self-Attention Module to obtain attention-guided features grabbing the global dependencies. In the block of the upsampling and downsampling processes, after passing through convolution or deconvolution layer to downsample or upsample, the fused features are processed by Dual Pooling-Attention Module to obtain attention-guided features. The motivation for this allocation comes from the different computational costs required by the two attention mechanisms. Dual Pooling-Attention Module is lighter, less expensive, and easier to implant into every part of the model, whereas Dual Self-Attention Module requires more computing resources, but provides a stronger ability to capture global dependencies. Both attention modules include spatial-based and channel-based parts, which can selectively aggregate contexts based on spatial attention maps, and at the same time emphasize class-dependent feature maps, helping to enhance the channel distinguishability of original features. Finally, a classifier with 1×1×1 convolution is used to classify the feature maps into target classes.

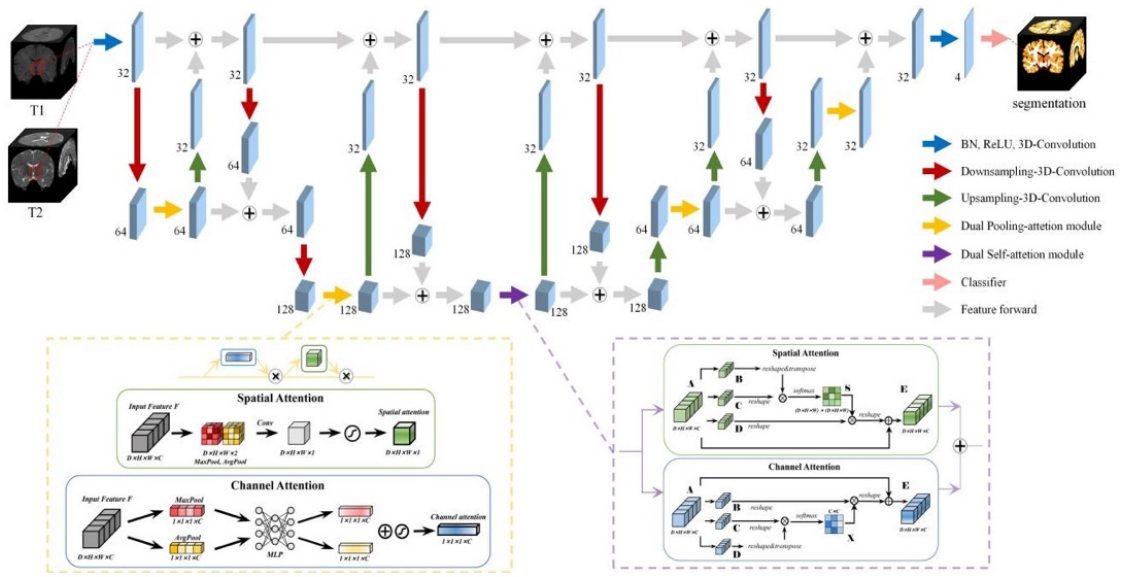

Fig. 5. Team *Tao_SMU*: proposed attention-guided full-resolution network architecture.

The training input of the network is 32×32×32×2 patches of T1w and T2w images. Both them are normalized to zero mean and unit variance. The network is trained with Adam optimizer on the cross-entropy loss, using minibatches of 4. During prediction, the step size is set as 4. It takes about 48 hours for training and 12 minutes for segmenting each subject on a TitanX Pascal GPU and Tensorflow framework.

### F. CU_SIAT: The Chinese University of Hong Kong and Shenzhen Institutes of Advanced Technology, China

To deal with the large differences between training images and testing images from multiple sites, Yu *et al*. adopt an Entropy Minimization Network (EMNet) to conduct unsupervised domain adaptation to align distributions between training and testing datasets.

As shown in Fig. 6, the proposed framework contains two main components: segmentation network and discriminator network. For the segmentation part, they employ the 3D densely connected network [16] to predict the brain structures by taking the T1w and T2w images as inputs. To enhance the generalization capability of the network over different site data, they carefully integrate Instance Normalization (IN) and BN as building blocks of the segmentation network [20]. The so called "IBN" layer design can improve network performance across different site data. Moreover, they adopt a discriminator network to perform adversarial entropy minimization [21] to align the distributions between training and testing datasets. In the problem setting, the training dataset is regarded as a source domain and testing dataset is treated as a target domain. They firstly compute the entropy maps of segmentation predictions for the source and target domain, respectively. And then, these entropy maps are fed into the discriminator network to produce domain classification outputs. The whole framework is trained with adversarial learning, so that the discriminator network is trained to distinguish the inputs from the source or target domain, while the segmentation network is trained to generate similar predictions for the source and target domain data to fool the discriminator network. Specifically, the optimization objective for segmentation network can be written as:

$$\min_{\theta s} \frac{1}{|\chi_s|} \sum_{x_s} L_{seg}(x_s, y_s) + \lambda_{adv} \frac{1}{|\chi_t|} \sum_{x_t} L_D\left(I_{x_t}, 1\right)$$

and the objective of discriminator network is

$$\min_{\theta_D} \frac{1}{|\chi_s|} \sum_{x_s} L_D\left(I_{x_s}, 1\right) + \frac{1}{|\chi_t|} \sum_{x_t} L_D\left(I_{x_t}, 0\right)$$

where $\theta_S$ and $\theta_D$ denote parameters of segmentation and discriminator network, respectively, $\chi_s$ and $\chi_t$ represent the set of source and target examples, $I_{x_s}$ and $I_{x_t}$ denote entropy maps of corresponding images. The $L_{seg}(x_s, y_s)$ and $L_D(I_{x_t}, 1)$ are cross-entropy losses and adversarial loss to train the segmentation network, and $\lambda_{adv}$ is used to balance the weights of the two losses.

The segmentation and discriminator network are all trained with Adam optimizer with a mini-batch size of 4. The learning rate is initially set as 2e-4 and decreased by a factor of $\beta = 0.1$ every 10000 iterators. They totally train 35000 iterations. The $\lambda_{adv}$ is set as 0.001. They use 10 labeled training samples and 16 unlabeled testing samples to train the whole framework. Due to the limited GPU memory, sub-volume of size 64×64×64 is used as network input. In the testing phase, they employ the sliding window strategy to generate the whole probability map of each test volume. Note that the discriminator network is abandoned during the testing phase. Generally, it takes about 24 hours to train the model and about 8 seconds to test one object on a GeForce RTX 2080Ti GPU with PyTorch library.

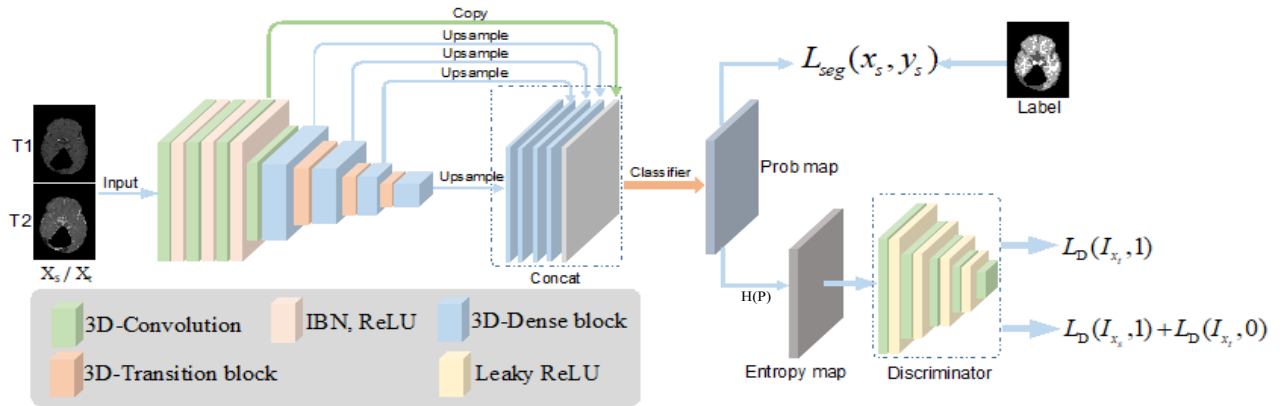

Fig. 6. Team *CU_SIAT*: the architecture of entropy minimization network (EMNet). The segmentation network is based on [16] with IBN layer. An adversarial entropy minimization strategy is used to minimize the self-information (i.e., entropy map) H(P) distribution gap between source and target domains.

### G. RB: Concordia University, Montreal, Canada

Based on the U-Net architecture [12], Basnet *et al*. propose a 3D deep convolutional neural network. The primary distinction between U-Net and the proposed architecture is the utilization of densely connected convolutional layers as building blocks of contracting the path and residual skip connections between contracting and expanding paths in the latter.

Fig. 7 shows the proposed network architecture. The contracting path begins with 3 sequential 3×3×3 convolutional layers with 32 kernels. The convolutional layer is followed by BN and a ReLU. Then, 4 downsampling dense blocks are stacked to halve the feature resolution at each block and gradually capture the contextual information. Each dense block has 8 convolution layers and starts with a max-pooling layer as shown in Fig. 8. In the dense block, every convolutional layer is preceded by a BN-ReLU. After each pair of convolutional layers in the block, a dropout layer with dropout rate of 0.2 is used to reduce overfitting. The output feature maps of each pair are concatenated with that of the previous pairs forming the dense connections. The first convolutional layer in the pair has 64 kernels of size 1×1×1 and the second one has 16 kernels of size 3×3×3. After every dense block, a 1×1×1 convolutional layer preceded by BN-ReLU called the transitional block is used and set to halve the number of feature maps. In the expanding path, 4 bilinear upsampling layers are used, that are set to double the feature resolution and halve the number of feature maps. A transitional block is used after each upsampling layer to match the number of feature maps to that of the transitional blocks in contracting path of the same resolution. The corresponding feature maps from the transitional blocks from contracting and expanding paths are elementwise summed using skip connections. At the end of fourth transitional block in the expanding path, the output features of the third convolutional layer from the beginning of the contracting path are added and fed to a 1×1×1 convolution layer preceded by BN-ReLU to get probability scores for the four output channels: WM, GM, CSF and background. The proposed network has 625,926 learnable parameters with 48 layers.

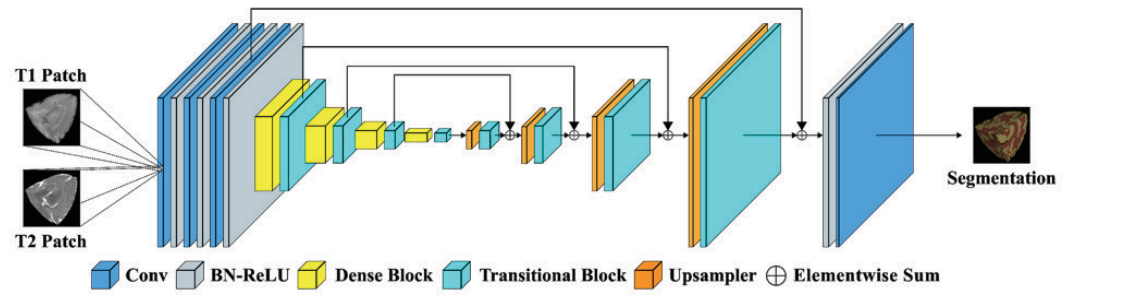

Fig. 7. Team *RB*: network architecture for U-DenseResNet.

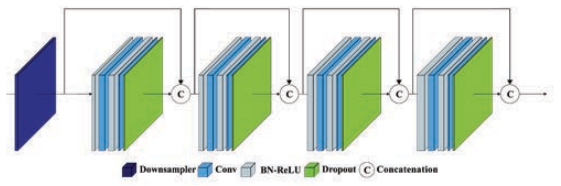

Fig. 8. Dense block used in Fig. 7.

For training, the randomly cropped 64×64×64 patches of normalized T1w and T2w scans to zero mean and unit variance are fed to the network. The network is trained on subjects 1-8 and 10 of iSeg-2019 dataset and the subject 9 is used for validation. The network is trained using Adam optimizer on a combination of cross-entropy and dice loss using minibatches of size 2 for 9600 epochs. The initial learning rate is set to 2e-4 and decreased by factor of 0.1 after 5000 epochs. The method is implemented using the PyTorch package on a computer with Intel Core i5-9600K CPU @ 3.70GHz, 16GB RAM and an NVIDIA RTX 2070 GPU. It takes approximately 16 hours to train the network. For inference, the T1w and T2w sequences cropped to patches of size 64×64×64 with overlapping steps of 16×16×16 are passed through the network and the probability scores at the overlapping regions are averaged to get the final segmentation results. The average inference time for segmenting a subject (256×256×256) is approximately 70 seconds.

### H. SmartDSP: Xiamen University and Chinese University of Hong Kong, China

Ma *et al*. propose a framework with adversarial learning of unsupervised cross-domain global feature alignment based on the backbone of nnU-Net [13].

The network employs nnU-Net [13] as the backbone, and the architecture of Ada-nnUNet model is shown in Fig.9. Inspired by [22] and [23], the network utilizes a feature-level domain discriminator at the bottleneck layer of nnU-Net, for aligning the distributions of target features and the source features in a compact space. In this challenge, they use the training dataset as source domain, and the unlabeled test dataset as target domain, thus formulating the problem as an unsupervised domain adaptation task. Specifically, the feature-level discriminator block consists of two 3×3 convolutional layers and one 1×1 convolutional layer. Then, the feature maps are forwarded to a global average pooling layer. This domain discriminator is trained to predict the domain of input features in a highly abstracted feature space, by practically setting the domain label as one for the source domain and zero for the target domain. By denoting $F_\theta$ as the feature extraction down-sampling layers before the bottleneck layer, $U_\phi$ as the up-sampling layers of the segmenter, and $D_\psi$ as the feature-level domain discriminator, the overall objective of the network is:

$$\max_{D_\psi} \min_{F_\theta, U_\phi} \mathcal{L}_{task}(F_\theta, U_\phi) + \mathcal{L}_{adv}(F_\theta, D_\psi)$$

where $\mathcal{L}_{task}$ is the task loss which is the sum of dice loss and cross-entropy loss in their method. For the loss of feature-level domain discriminator, they expect the classifier to alleviate the imbalance of alignment difficulties across different samples, they follow [23] and apply the focal loss for $\mathcal{L}_{adv}$ as:

$$\mathcal{L}_{adv} = -\frac{1}{N_s}\sum_{i=1}^{N_s}\left(1 - D_\psi\left(F_\theta(x_s^i)\right)\right)^\gamma \log\left(D_\psi\left(F_\theta(x_s^i)\right)\right)$$
$$= -\frac{1}{N_t}\sum_{i=1}^{N_t}\left(D_\psi\left(F_\theta(x_t^i)\right)\right)^\gamma \log\left(1 - D_\psi\left(F_\theta(x_t^i)\right)\right)$$

where $N_s$ denotes the number of source examples and $N_t$ denotes the number of target examples, and each training sample drawn from source domain and target domain is represented by $x_s^i$ and $x_t^i$, respectively. The $\gamma$ adjusts the weight on hard-to-classify examples and was empirically set as 5.0 in the network. In each training batch, source image $x_s^i$ is also forwarded to minimize the segmentation network for $\mathcal{L}_{task}$.

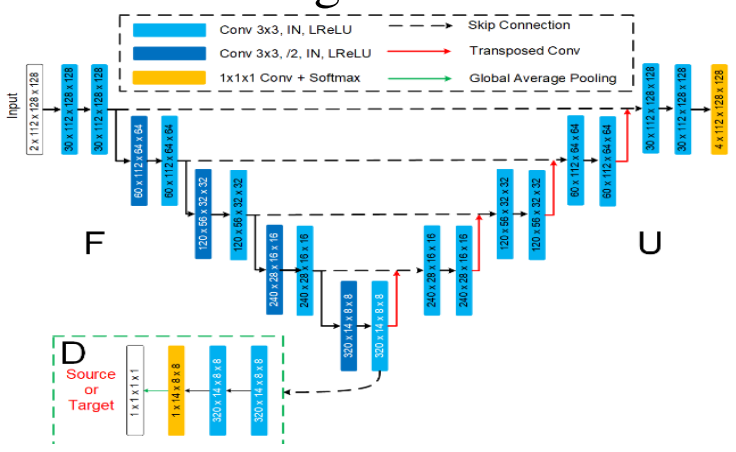

Fig. 9. Team *SmartDSP*: the proposed Ada-nnUnet network.

The network is implemented by using the PyTorch toolbox, with Adam optimizer. The initial learning rate is set as 3e–4 and the weight decay regularizer is set as 3e–5. The inputs of the network contain T1w and T2w MR images and there is no extra training dataset used. In the training stage, patches with size of 112×128×128 are randomly extracted from the volumes and input to the network. The training procedure takes around two days on a GPU of NVIDIA TITAN Xp with 12 GB memory and it takes around 10 seconds to process one subject during the testing phase. The potential limitation of Ada-nnUNet lies in that their feature-level alignment is imposed only in the coarse feature space. In this regard, some detailed information would have to be neglected during adversarial distribution alignment, which may impede segmentation performance. In future work, they will elaborate the feature alignment into multiple levels and spatial attentions.

## IV. Results and Discussion

In this section, Dice coefficient (DICE), 95th-percentile Hausdorff distance (HD95), and average surface distance (ASD) [1] are adopted to evaluate the performance of the 8 top-ranked methods. First, segmentation results of different teams are presented in Fig. 10. Evaluations in terms of the whole brain using DICE, HD95 and ASD are presented in Fig. 11 and Appendix Table III respectively. For simplification, {WM, GM, CSF}-{DICE, HD95, ASD} denotes the performance on the soft tissue (WM, GM and CSF) in terms of a metric (DICE, HD95 and ASD). Besides evaluations for the whole brain, we also evaluate the performance based on small regions of interest (ROIs), gyral landmark curves and cortical thickness in Figs. 12-14, Appendix Table VII and VIII. In order to compare the difference of segmentation results among the 8 top-ranked methods, Wilcoxon signed-rank tests are calculated for statistical analysis in Appendix Table IV, V and VIII. Furthermore, Wilcoxon rank-sum tests are applied to analyze the statistically significant difference among multiple sites/teams in Appendix Table VI.

First, Fig. 10 qualitatively shows the segmentation results of different teams, and Appendix Table III quantitively listed the performances evaluated by using DICE, HD95 and ASD. To better demonstrate the performances by different teams, Fig. 11 further employs violin-plots to illustrate the performance distribution of each testing subject. The shape of violin-plots reflects the distribution of evaluation results, that is, the wider (shorter) the shape, the more stable the performance of methods. Obviously, from Fig. 10, these teams consensually have a relatively better performance on testing subjects from UNC/UMN (BCP) and Emory University sites, compared with Stanford University site. Then from the quantitative analysis in Appendix Table III, 7 out of 8 top-ranked methods have relatively better segmentation results on UNC/UMN (BCP) site in terms of three metrics, which can be also observed directly from Fig. 11. In Fig. 11, the light blue, orange, grey points, denote the performances on the testing subjects from UNC/UMN (BCP), Stanford University, and Emory University, respectively. It can be seen the light blue points are always at the top location with regard to DICE metric in the first column, while the orange points and grey points are located in the bottom and middle, respectively. As for HDF5 and ASD, the corresponding point location is opposite as shown in the second and third columns. We can conclude from the above analysis, in terms of the whole brain, most methods achieved the best performance on testing subjects from UNC/UMN (BCP) site and the worst performance on those from Stanford University, while in between on those from Emory University.

Besides, the violin-plot shapes of *QL111111*, *FightAutism* and *trung* teams are much wider, which indicates a more stable performance on the testing subjects from three sites than other teams, as shown in Fig. 11. Furthermore, in order to compare the significant difference of segmentation results among the 8 top-ranked methods, Wilcoxon signed-rank tests are calculated as illustrated in Appendix Table IV (on three sites totally) and V (on three sites separately) with all-against-all diagram in terms of three metrics (i.e., DICE, HD95 and ASD). We can find that only *QL111111* have strongly statistically significant difference on WM-DICE and WM-ASD compared with other teams as illustrated in Table IV ($p$-value < 0.01). However, from the evaluation on three sites separately reported in Appendix Table V, *QL111111* only has weak statistical significance on WM-ASD when testing on UNC/UMN (BCP) site, and on WM-ASD when testing on Stanford University site ($p$-value < 0.05). As for HD95, there is no strong, statistically significant difference among these teams neither for evaluation of all testing sites nor single testing site in Appendix Table IV and V respectively ($p$-value > 0.05). As for the ASD metric, *QL111111* has statistically significant differences in terms of WM and GM with other teams on all three sites as shown in Table IV ($p$-value < 0.01), whereas from the multi-site analysis reported in Table V, it only has weak statistical significance on GM-ASD when testing on Emory University site ($p$-value < 0.05).

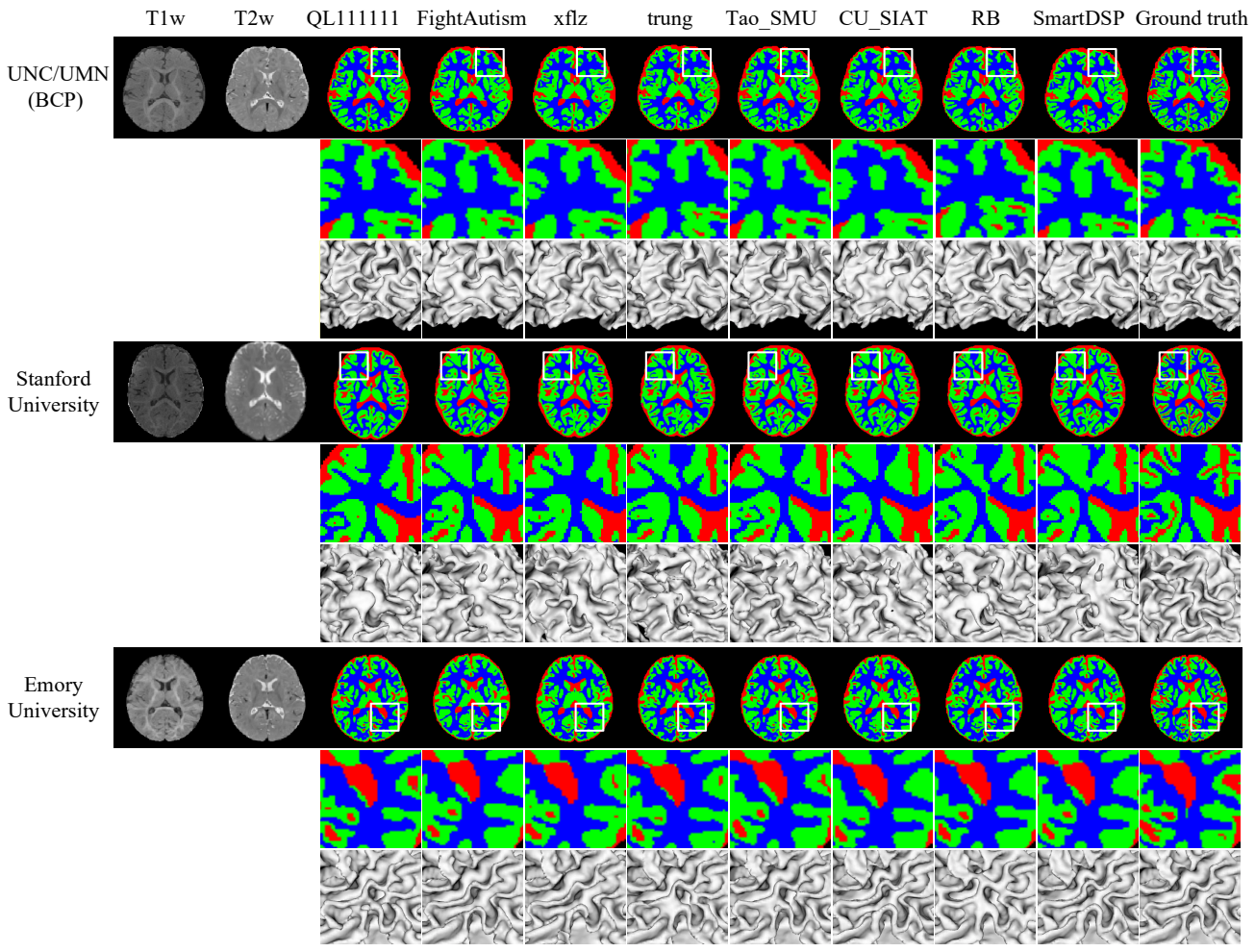

Fig. 10. Segmentation results by different teams on testing images from three sites. From left to right: T1w and T2w images, tissue segmentation results obtained by 8 top-ranked methods, and ground truth. Zoomed 2D segmentation results and corresponding 3D rendering results are also included.

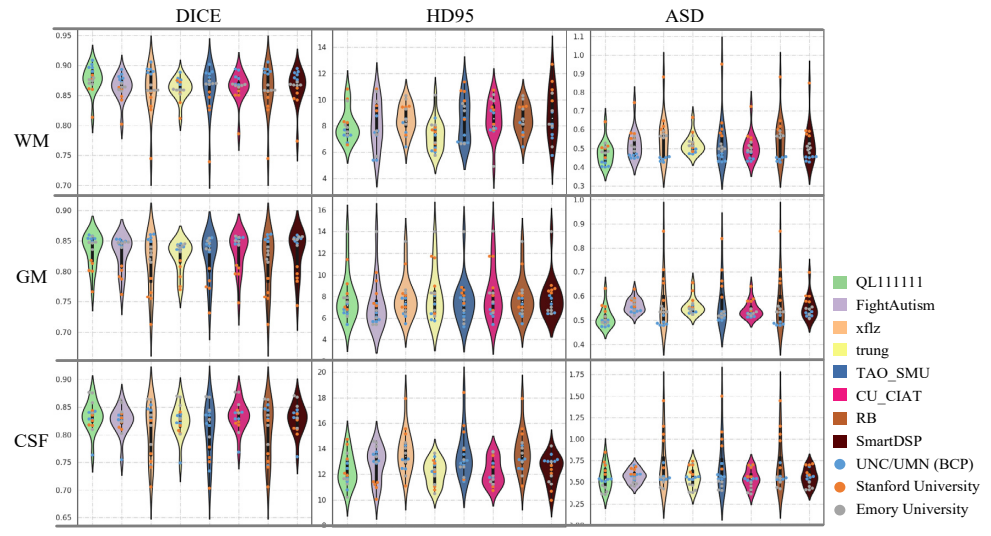

Fig. 11. Performance of the 8 top-ranked methods on tissue segmentation, in terms of DICE, HDF5 and ASD, using violin-plots. Testing subjects from three sites are marked as points in three colors.

Second, to further compare the performance of the 8 top-ranked methods in respect of small ROIs, 80 ROIs-based evaluation is employed in this paper. Following our previous review article in iSeg-2017 [1], we use a multi-atlas-based technique to parcellate each testing subject into 80 ROIs. In particular, a total of 33 two-year-old subjects from www.brain-development.org were employed as individual atlases, in which each case consists of a T1w MR image and the corresponding label image with 80 ROIs (excluding cerebellum and brainstem). First, each T1w MR image is segmented into WM, GM, and CSF tissues by iBEAT V2.0 Cloud (http://www.ibeat.cloud). Then, based upon the tissue segmentation maps, all anatomical atlases are warped into the space of each testing subject via ANTs [24]. Finally, each testing subject is parcellated into 80 ROIs using majority voting. Due to the large number of ROIs, we only employ DICE metric to measure the similarity of automatic segmentations with the manual segmentation as listed in Appendix Table VII. In addition, we use the violin-plot in Fig. 12(a) to demonstrate the distribution of DICE values among three testing sites, where the different color points in violin-plot indicate different testing sites. As shown in Fig. 12(a), the DICE values from the Stanford University site show a larger distribution range (the length of violin-plot) compared with other two sites, which indicates segmentation performance of the methods on this site is unstable. Furthermore, Fig. 12(b) calculates the mean DICE values of each site to further show the performance on each site, which is consistent with Fig. 10 and Fig. 11, that is, all of 8 top-ranked methods perform poorly on the Stanford University site due to its different imaging protocol/scanner.

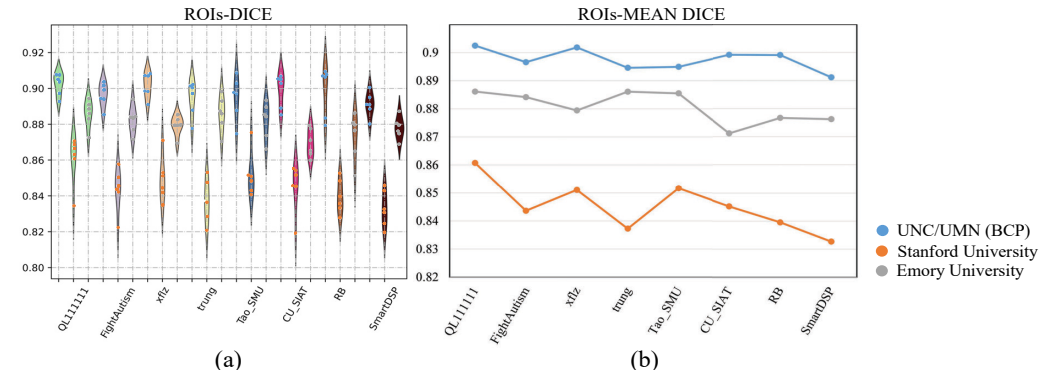

Fig. 12. Performance of the 8 top-ranked methods on ROI-based evaluation. (a) The DICE values of each team on three sites. For the violin-plots, different color points indicate different testing sites. (b) The mean DICE values of each team on three sites.

Third, in human brains, gyri are part of a system of folds and ridges in the cerebral cortex that create a larger surface area, and changes in the structure of gyri are associated with various diseases and disorders [25]. Moreover, the cortical thickness estimation is highly sensitive with the segmentation accuracy [9]. Therefore, in this subsection, we further measure the distance of gyral landmark curves on the cortical surfaces, as well as the cortical thickness. In detail, for the gyral curves, we first reconstruct the inner cortical surface using the in-house cortical surface reconstruction pipeline [26] for each segmentation. Then, the typical gyri anchor points for the gyri are manually marked according to the mean curvature pattern of the reconstructed inner cortical surface with the paraview toolkit (https://www.paraview.org). Specifically, these anchor points are selected as the local maximum of the mean curvature on the corresponding gyri. For two neighboring gyral anchor points from the same gyri on the surface, we connect them with the minimal geodesic distance path; and finally, the entire gyral curve can be obtained by connecting all the gyral anchor points. For the cortical thickness maps, based on the inner and outer cortical surface reconstructed, the paraview toolkit is used to generate the thickness maps.

Fig. 13 shows two major gyri (i.e., the superior temporal gyral curve and the postcentral gyral curve) and the cortical thickness maps. From top to bottom, there are three subjects randomly selected from three sites respectively. For example, in the first row for UNC/UMN (BCP) site, we can observe that these methods perform relatively accurate gyral landmarks, and cortical thickness is within a normal range. Then, for the second row, i.e., Stanford University site, the gyral curves of methods are quite different with each other and unsmooth. Consequently, the corresponding cortical thickness are abnormally thicker or thinner.

Additionally, HD95 metric is also employed to calculate the curve distance between the estimated landmarks with the ground truth, as reported in Fig. 14. A large curve distance indicates poor performance on the gyral crest. As shown in Fig. 14(a) and (b), these methods consistently have higher HD95 values on the Stanford University site, which indicates a poor performance on the Stanford University site. Based on Wilcoxon signed-rank test, the $p$-values are calculated to evaluate the statistically significant difference, as shown in Appendix Table VIII. However, there is no statistically significant difference among these teams ($p$-value > 0.05).

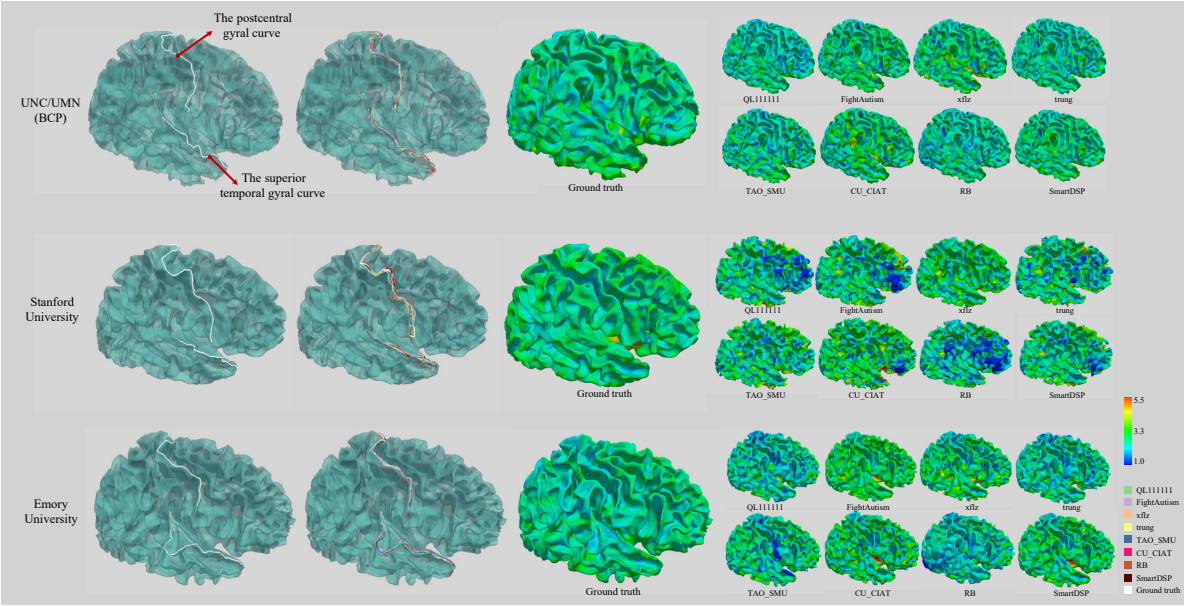

Fig. 13. Two gyral curves (i.e., the superior temporal gyral curve and the postcentral gyral curve) and cortical thickness maps for 8 top-ranked methods and ground truth on three sites respectively. From left to right: ground truth of two gyral curves, the gyral curves from the segmentation results of 8 top-ranked methods, cortical thickness maps of ground truth, and cortical thickness maps from the segmentation results of 8 top-ranked methods.

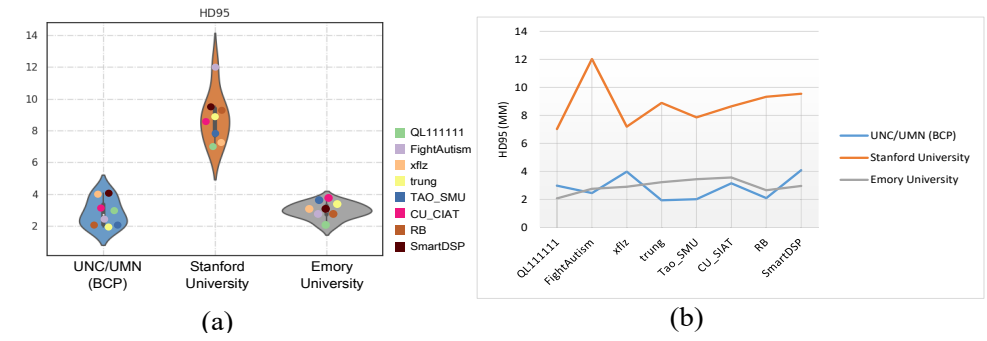

Fig. 14. HD95 evaluation of 8 top-ranked methods on the superior temporal gyral curve and the postcentral gyral curve. (a) Violin-plot shows HD95 evaluation distribution of three sites. (b) Line chart shows the difference of HD95 evaluation among three sites.

## V. Future Directions

Based on the above evaluation and discussion, regarding the whole brain, ROIs, gyral curves and cortical thickness, we can conclude that unfortunately none of these 8 top-ranked methods can handle the multiple-site issue, especially when the imaging protocols/scanners are different at the training and testing sites. The violin-plot shown in Fig. 15(a) illustrates the difference between validation and testing datasets. In Fig. 15(a), we can find there are two peaks in the violin-plots with regard to WM and GM, corresponding to the validation and testing datasets. In addition, the testing results from Stanford University are always at the bottom of each violin-plot, indicating a poor performance on this site. In addition, the Wilcoxon signed-rank tests among validation dataset and three testing datasets reported in Appendix Table VI also indicate a statistically significant difference for GM on the Stanford University site ($p$-value < 0.05), compared with other sites. Besides, Fig. 15(b) shows the DICE values of the 8 top-ranked methods for WM. Most methods cannot perform well on three testing sites, especially for the site from Stanford University, which is caused by the great difference in imaging parameters/scanner (i.e., Stanford University: GE scanner; Others: Siemens scanners). It can be observed that most of the teams cannot achieve a consistent performance on different sites, let alone the large performance drop from the validation dataset to testing datasets, even though some of them designed special methods to deal with the multi-site issue. Regarding these methods, *QL111111* used 3D U-Net combined with attention mechanism and dilated convolution to further capture multi-scale information and encode wider context information, however they only perform some common image preprocessing steps, like adjusting image contrast, to deal with the multi-site issue. *xflz* applied a tissue-dependent intensity augmentation method to simulate the variations in contrast, whereas it does not work well on the multi-site datasets. In order to decrease the distribution difference of multi-sites, *CU_SIAT* adopted EMNet to align distributions between training and testing datasets by using a discriminator network, and *SmartDSP* proposed Ada-nnUNet by utilizing a feature-level domain discriminator at the bottleneck layer of nnU-Net, but their methods did not achieve the best results nonetheless. It is worth investigating the essential data variance among different medical sites and how we can quantitatively measure them. For example, as suggested by *xflz* team, we could use imaging parameters to map the images from the testing site to the space of training images based on MR imaging physics.

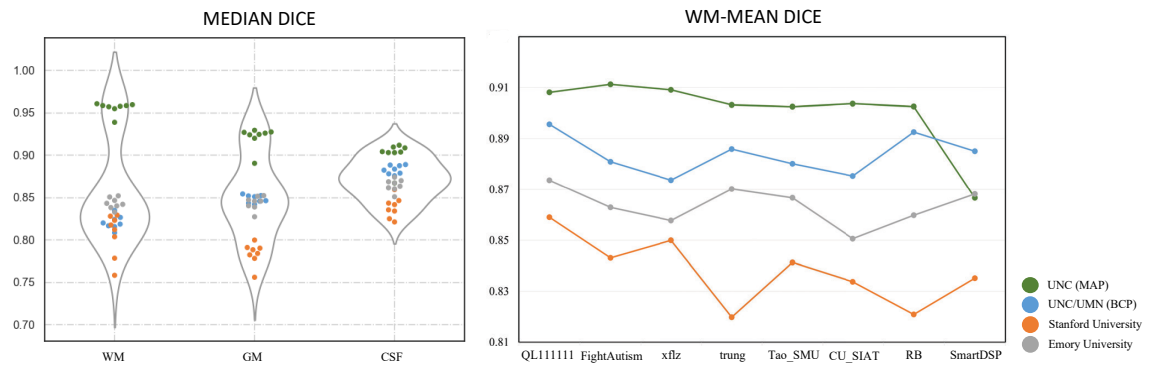

Fig. 15. (a) Average performance of 8 top-ranked methods on the validation subjects and testing subjects from 3 sites based on median DICE for WM, GM and CSF. (b) Performance of each 8 top-ranked method on WM-mean DICE.

Second, those reviewed methods in iSeg-2019 still did not consider any prior domain knowledge of infant brain images, e.g., cortical thickness is within a certain range [9, 26]. It is also well known that tissue contrast between GM and WM is extremely low due to inherent myelination and maturation process. However, all methods ignored the relatively high contrast between CSF and GM. The prior knowledge is site-independent/scanner-independent, which might be a key direction to explore to further deal with the multi-site issue task. Our previous works have demonstrated the effectiveness of using the prior in helping the tissue segmentation [9, 10].

Third, the key highlights and implementation details of the 8 top-ranked methods are listed in the Appendix Table VX. For example, all methods randomly selected 3D patches during the training stage, which could be improved by selecting patches from the error-prone regions. As shown in Fig. 16, the main error regions exist in cortical regions, such as straight gyrus and lingual gyrus, thus, more training patches from these regions may improve the segmentation performance. In addition, although training labels are always limited, we could employ a semi-supervised learning strategy or use existing methods to generate auxiliary labels [27] from the unlabeled testing subjects for a better training.

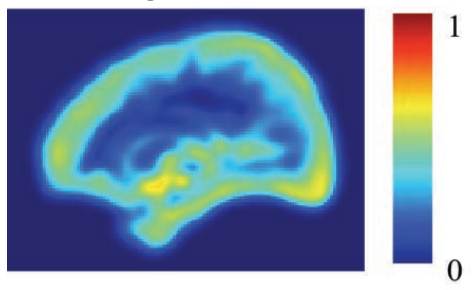

Fig. 16. Average error map for all 8 top-ranked methods. Color bar is from 0 to 1, with the high values indicating large errors.

Finally, we would like to indicate limitations for iSeg-2019.

First, due to word count limitations, only 8 top-ranked methods were reviewed in this paper, but some other teams demonstrate useful strategies. For example, *WorldSeg* added a contour regression block to cope with the blurry boundary problem. *SISE* used 3D CycleGAN for domain adaption between different sites. *SJTU-IMR* computed distance maps acquired by 3D U-Net to model the spatial context information, which can be viewed as one channel of FCN to get the final segmentation. *MASI* applied the pertained model based on adult SLANT on the testing subjects. Second, only three sites are included in iSeg-2019 and for each site only 5 or 6 subjects are provided. Third, only Siemens and GE scanners are included in this challenge. For our planed future challenge, we will include more sites with various scanners/models and more testing subjects.

## VI. Conclusion

In this paper, we reviewed and summarized 30 automatic infant brain segmentation methods participated in iSeg-2019 involving multi-site imaging data. In particular, we elaborated on the details of 8 top-ranked methods, including the pipeline, implementation, experimental results and evaluations. We also discussed their limitations and possible future directions. The multi-site issue is still an open question and the iSeg-2019 website is always open. We hope our manual labels in iSeg-2019 and this review article will help to advance methodological development on the multi-site issue.